\documentclass[aps,pre,twocolumn,superscriptaddress,floatfix]{revtex4}
\usepackage[english]{babel}
\usepackage{graphicx,color}
\usepackage[dvips,unicode,colorlinks,linkcolor=blue,citecolor=blue,urlcolor=blue]{hyperref}
\usepackage{times}

\begin{document}
\title{Confinement and non-universality of anomalous heat transport and  superdiffusion of energy in low-dimensional systems}
\author{Yuriy A. Kosevich}
\email[]{yukosevich@gmail.com}
\affiliation{Semenov Institute of Chemical Physics, Russian Academy of Sciences,
Moscow 119991, Russia}
\author{Alexander V. Savin}
\email[]{asavin@center.chph.ras.ru}
\affiliation{Semenov Institute of Chemical Physics, Russian Academy of Sciences,
Moscow 119991, Russia}
\date{\today}
\begin{abstract}
We provide molecular dynamics simulation of heat transport and thermal energy diffusion in
one-dimensional molecular chains
with different interparticle pair potentials at zero and non-zero
temperature.  We model the thermal conductivity (TC) and energy diffusion
in the coupled rotator chain and in the Lennard-Jones chain either without or with the confining
parabolic interatomic potential. The considered chains without the confining potential have normal TC and energy diffusion,
while the corresponding chains with the confining potential are characterized by anomalous
(diverging with the system length) TC and superdiffusion of energy.  We confirm in such a way that, surprisingly,
the confinement makes both heat transport and energy diffusion anomalous in low-dimensional phononic
systems. We show that the chain, which has a finite TC, is also
characterized by the normal energy diffusion in the thermalized chain (at non-zero temperature),
while the superdiffusion of thermal energy occurs in the thermalized chains with only anomalous TC.
We present the arguments, supported by our simulations, that the scaling relation between
the exponents in time dependence of the mean square displacement of thermal energy distribution
and in length dependence of the anomalous TC is not universal and can be different,
depending on the main mechanism of energy transport: by weakly-scattered waves
or by noninteracting colliding particles performing L\'{e}vy flights.
\end{abstract}
\maketitle

Phonon thermal conductivity in low-dimensional systems and at nanoscale presents a challenge
both from theoretical and experimental points of view \cite{reviewI,reviewII}.
One of important problems in this field is the origin of anomalous, diverging with the system size,
thermal conductivity (TC) of one-dimensional (1D) lattice models \cite{llp97,h99,nr02,llp03,cp03,lw03,dllp06,dhar08}
and quasi-1D nanostructures and polymers \cite{hc08,yzl10,ly12}.
The length-dependent TC was experimentally observed in carbon nanotubes \cite{cz08}
and in suspended single-layer graphene \cite{x14}. The
general belief is that TC is anomalous in 1D momentum-conserving systems \cite{pc00}.
Nevertheless, a chain of coupled classical rotators presents an example of
translationally-invariant (isolated) and correspondingly
momentum-conserving 1D periodic system with finite TC \cite{glpv00,gs00}.
In a recent paper \cite{sk14}, normal heat transport
was reported for 1D momentum-conserving systems with the
Lennard-Jones, Morse, and Coulomb potential. It was shown in \cite{sk14} that the convergence
of TC  is provided by phonon scattering on the locally strongly stretched loose interatomic bonds
at low temperature and by the many-particle scattering at high temperature.

While the research on TC in low-dimensional systems is very intensive,
the peculiarities of energy diffusion have been much less studied.
Usually the nonthermalized chain has been considered with only small central
portion being thermalized, and thermal energy propagation in the chain was studied afterwards.
Such modeling can lead to anomalous (super-) diffusion of the thermal energy, and sometimes
it is concluded that such 1D system possesses anomalous thermal conductivity \cite{lllhl14}.
For example, superdiffusion of energy was obtained in Ref. \cite{lllhl14} in the  chain with
the Lennard-Jones (LJ) interparticle potential, in which the normal heat transport has been
revealed previously \cite{czwz12,sk14}. This suggests the incorrectness of the simulations
of energy diffusion because the latter was simulated in Ref. \cite{lllhl14} in the chain
at zero background temperature. Any anharmonic chain at low temperature behaves as
an almost harmonic chain, in which the heat transport is carried out by ballistic phonons
and TC diverges with the chain length. Therefore it is not surprising that in the
nonthermalized chain (at zero background temperature) the energy propagates ballistically.
But this does not imply that the superdiffusion of energy will also hold in the thermalized
chain (at non-zero background temperature).

On the other hand, the chains with a (parabolic or quartic) confining pair potential, which
does not allow for bond dissociation, are expected to possess anomalous thermal conductivity,
diverging with the chain length \cite{sk14}. The celebrated Fermi-Pasta-Ulam potential belongs
to such type of the pair potentials.
In this paper, we model both the TC and diffusion of energy in anharmonic chains at zero and non-zero
temperature. We show that the chain, which has a finite TC, is also characterized by the
normal diffusion of energy in the thermalized chain, while the anomalous superdiffusion of thermal
energy occurs in the thermalized chains with only anomalous (diverging with the system length) TC.
To show this, we model TC and energy diffusion
in the coupled rotator chain (a chain with periodic interatomic potential) and in LJ chain
either without or with the parabolic confining pair potential. Such parabolic confining pair
potential is introduced, for instance, on top of the shallow LJ  potential  to
hold together the constituent cellular particles within a cell \cite{kosztin12}.
The considered chains without
the confining potential are characterized by the
normal diffusion of energy in the thermalized chain, while the corresponding chains with the confining pair potential are characterized by the superdiffusion of energy.
We confirm in such a way that, surprisingly, the confinement makes both heat transport
and energy diffusion anomalous in low-dimensional systems. We present the
arguments, supported by our simulations, that the scaling relation between the exponents in time dependence
of the mean square displacement of thermal energy distribution and in length dependence of the TC,
which is discussed in connection with anomalous heat transport and superdiffusion of energy in low-dimensional
systems, see, e.g., Refs. \cite{l05,den2003,cdp05,dhar2013,li2014},
is not universal and can be different, depending on the main mechanism
of energy transport -- either by weakly-scattered waves
or by noninteracting colliding particles performing L\'{e}vy flights.

We consider the periodic molecular chain consisting of $N=L/a$ unit cells, where $L$ is the chain length, $a$ is a lattice period. In
a dimensionless form, the Hamiltonian of the chain can be written as
\begin{equation}
H=\sum_{n=0}^N\frac12\dot{x}_n^2+\sum_{n=0}^{N-1}V(x_{n+1}-x_n), \label{f1}
\end{equation}
where $x_n$ is the displacement on the $n$th particle from its equilibrium position at $na$,
$V(r)$ is a dimensionless pair interaction potential between nearest neighbors normalized by the conditions
$V(0)=0$ and $V'(0)=0$.

To simulate the heat transfer in the chain, we use the
stochastic Langevin thermostat. We consider a finite chain of
$N_+ +N+ N_-$ unit cells, and take the chain with fixed ends.
We put the $N_-$ left boundary particles in the Langevin thermostat with temperature $T_-$,
and $N_+$ particles in the right-hand edge Langevin thermostat with a temperature of $T_+$.
The corresponding system of equations of motion of the chain is:
\begin{eqnarray}
\ddot{x}_n&=&-\partial H/\partial x_n-\gamma\dot{x}_n+\xi^+_n,~~n\le N_+, \nonumber\\
\ddot{x}_n&=&-\partial H/\partial x_n,~~n=N_++1,...N_++N, \label{f2} \\
\ddot{x}_n&=&-\partial H/\partial x_n-\gamma\dot{x}_n+\xi^-_n,~~n> N_++N, \nonumber
\end{eqnarray}
where $\gamma$ is relaxation coefficient of the particle velocity, $\xi_n^\pm$ are random forces which
simulate the interaction with the white-noise thermostat normalized
by the conditions
$$
\langle\xi^\pm_n(t)\rangle=0,~~
\langle\xi_n^\pm(t_1)\xi_k^\pm(t_2)\rangle=2\gamma T_\pm\delta_{nk}\delta(t_2-t_1).
$$

We integrate numerically the Langevin equations of motion (\ref{f2}) by employing the Verlet
velocity method with the step $\Delta t=0.02$.
After some integration time $t_0$ (this value depends on the chain length between the thermostats),
we observe the formation of a temperature gradient and constat heat energy flux in the central
part of the chain. After the stationary heat flux is established, we can find the temperature distribution
by using the relations for $T_n =\langle\dot{x}_n^2\rangle_t$ and stationary heat flow
along the chain $J_n=-\langle\dot{x}_nV'(x_n-x_{n-1})\rangle_t$. The
following values were used in the numerical simulation: $T_\pm=(1\pm 0.1)T$,
$\gamma=0.1$, $N_\pm=40$, $N=20$, 40, 80, ..., 20480.
The detailed description of the Langevin equations and
justification of the method are presented in \cite{sk14}.

In the steady-state regime, the heat flux through each cell in the central part of the chain
should be the same, i.e., $J_n\equiv J$, $N_- + 1 \leq n \leq N_- + N + 1$.
Almost linear gradient of temperature distribution is established in the central part of the chain,
so we can define TC as
\begin{equation}
\kappa(N)=J(N-1)/(T_{N_++1}-T_{N_++N}). \label{f3}
\end{equation}

In our modeling of the TC, we use the following pair interaction potentials:
\begin{equation}
V(r)=1-\cos(r),~~a=2\pi, \label{f4}
\end{equation}
-- the periodic potential,
\begin{equation}
V(r)=1-\cos(r)+0.25r^2,~~a=2\pi,  \label{f5}
\end{equation}
-- the sum of the periodic and parabolic confining potentials,
\begin{equation}
V(r)=4\epsilon [(\sigma/(1+r))^6-1/2]^2,a=1,\sigma=2^{-1/6},\epsilon=1/72, \label{f6}
\end{equation}
-- the LJ potential,
\begin{equation}
V(r)=4\epsilon [(\sigma/(1+r))^6-1/2]^2+0.25 r^2,  \label{f7}
\end{equation}
-- the sum of the LJ and parabolic confining potentials. Note that  one has $V''(0)=1$ for potentials (\ref{f4})
and (\ref{f6}) and therefore speed of sound (small-amplitude phonons) in the chain is $v_s=a$, while
for potentials (\ref{f5}) and (\ref{f7}) one has $V''(0)=1.5$ and therefore $v_s=a\sqrt{1.5}=1.2247a$.
\begin{figure}[tb]
\begin{center}
\includegraphics[angle=0, width=1\linewidth]{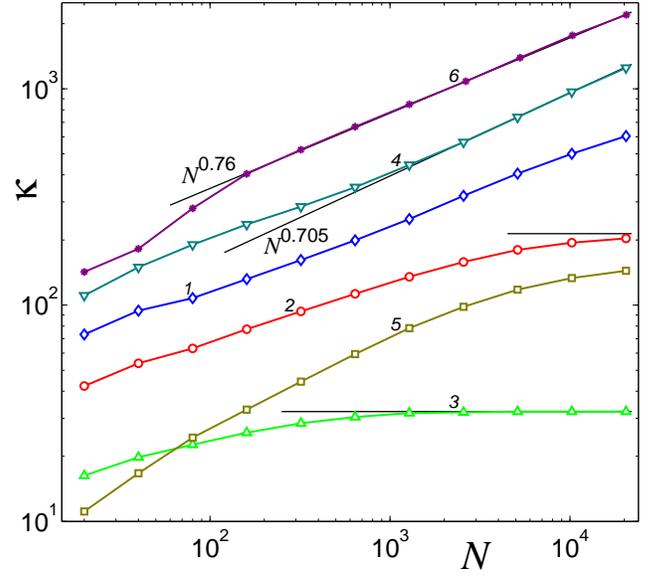}
\end{center}
\caption{
Thermal conductivity $\kappa$ versus dimensionless chain length $N=L/a$ for the chain with the periodic
potential (\ref{f4}) at normalized temperature $T=0.15$, 0.2, 0.3
(curves 1, 2, 3), for the chain with the combined potential (\ref{f5}) at $T=0.3$ (curve 4),
for the chain  at $T=0.002$ with the Lennard-Jones potential (\ref{f6})
(curve 5, which shows for the convenience $0.15\kappa$) or with the combined potential (\ref{f7}) (curve 6).
}
\label{fig01}
\end{figure}

Dependence of TC on dimensionless chain length $\kappa(N)$ is presented in Fig.~\ref{fig01}. As one can see from
this figure,
$\kappa(N)$ in the chain with the periodic potential (4) depends
on temperature: the convergence of $\kappa(N)$ for $N\rightarrow\infty$ is slower for lower temperature.
Thermal conductivity saturates for $N=2560$ at $T=0.3$, for $N=20480$ at $T=0.2$, and there is
no saturation of the $\kappa(N)$ at the maximal used $N=20480$  at $T=0.1$
(and one needs $N\sim10^7$ to reach the saturation at such temperature).
This feature is related with the increase of phonon mean free path with the decrease
of temperature because the saturation of $\kappa(N)$ occurs only when the chain length $L=Na$ reaches or exceeds  phonon mean free path.

The chain with the combined interatomic potentials (\ref{f5}) and (\ref{f7})
is characterized by anomalous heat transport.
Here TC monotonously diverges with $N$ as  $\kappa(N)\sim N^\alpha$. For the potential (\ref{f5})
at $T=0.3$, the exponent $\alpha=0.705$, for the potential (\ref{f7}) at $T=0.002$, the exponent $\alpha=0.76$.
TC of the  chain with the LJ potential (\ref{f6}) saturates for $N\sim 10^4$ at $T=0.002$, as one can see in Fig. \ref{fig01}.

Now we analyze how the $\kappa(N)$ dependence is related with thermal energy propagation
at finite temperature ($T_0>0$). To this end, we consider the chain with $N$ atoms with the fixed ends.
We thermalize the chain with the use of the Langevin heat bath such that the first
$N_l$ atoms have the temperature $T_l$, while the rest of the atoms have the temperature $T_0<T_l$.
We consider the Langevin equations of motion with random forces:

\begin{equation}
\ddot{x}_n=-\partial H/\partial x_n-\gamma\dot{x}_n+\xi_n,~~0<n<N, \label{f8}
\end{equation}
when $x_0\equiv 0$, $x_N\equiv 0$, and the random forces $\xi_n$ are $\delta$-correlated as
$$
\langle\xi_n(t_1)\xi_k(t_2)\rangle=2\gamma T_n\delta_{nk}\delta(t_2-t_1),
$$
with $T_n=T_l$ for $n=1,...,N_l$, and $T_n=T_0$ for $n=N_l+1,...,N-1$.
Now we take the (classical) zero-energy initial condition
$\{x_n(0)=0,~\dot{x}_n(0)=0\}_{n=1}^{N-1}$ and integrate the Langevin equations (\ref{f8})
from $t=0$ to $t_1=20/\gamma=200$. In result we get a random realization of the initial
state of the thermalized chain $\{x_n(t_1),\dot{x}_n(t_1)\}_{n=1}^{N-1}$
with $T=T_l$ in its left end and $t=T_0<T_l$ in the rest of the chain.
\begin{figure}[tb]
\begin{center}
\includegraphics[angle=0, width=1.\linewidth]{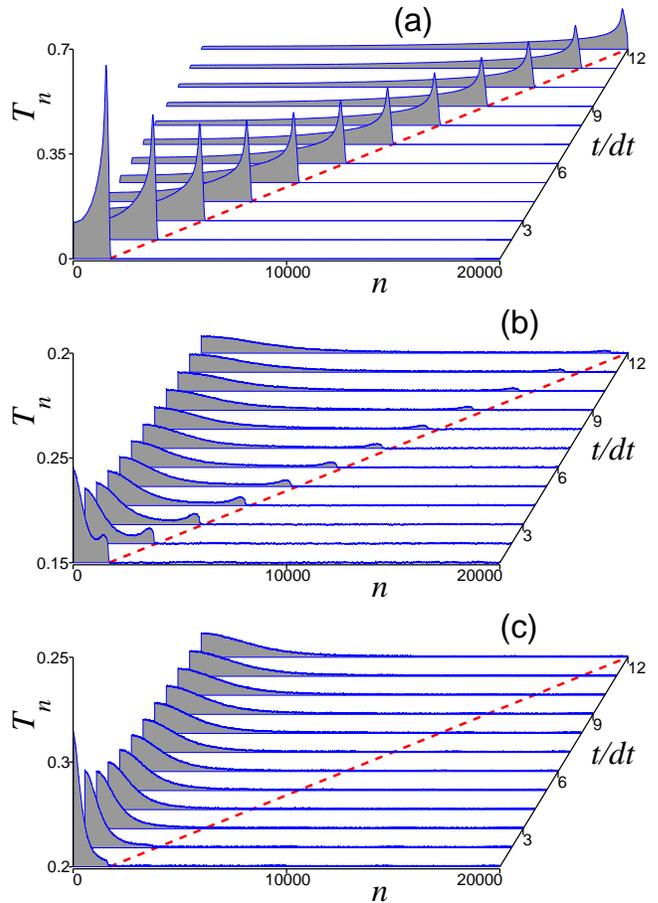}
\end{center}
\caption{
Propagation of temperature distribution $T_n$ in a chain with the periodic potential
(\ref{f4}) for $T_l=1$ and $T_0=0$ (a), $T_0=0.15$ (b), $T_0=0.2$ (c).
Temperature distributions are shown for characteristic delay time $dt=1662.5$,
dashed lines show sound cone $n=v_st/a$.
}
\label{fig02}
\end{figure}

Then we model the propagation of the thermal energy in the chain, which starts from
its left end with higher temperature. To perform this, we integrate equations of motion of the atoms
without their interaction with the thermostat,
\begin{equation}
\ddot{x}_n=-\partial H/\partial x_n,~~0<n<N, \label{f9}
\end{equation}
when $x_0\equiv 0$, $x_N\equiv 0$, with the initial condition which were obtained after
the integration of the Langevin equations of motion (\ref{f8}).  In order to analyze the
propagation of the thermal energy during the long time in the chain with $N=20000$ atoms,
we take the number of the end atoms $N_l=40\ll N$. We trace the time dependence of the distribution
 along the chain of the temperature  $T_n(t)=\langle\dot{x}_n^2(t)\rangle$
and of the energy $E_n(t)=\langle\dot{x}_n^2(t)/2+V(x_n(t)-x_{n-1}(t))\rangle$, where the averaging
is taken over the independent realizations of the initial thermalized state of the chain.
Time evolution of the temperature distribution in the chain with the periodic potential
is shown in Fig. \ref{fig02}. It is worth noting that for the analysis of thermal energy
distribution in the thermalized chain, with $T_0>0$, one needs to perform the averaging over
a large number ($\sim 10^5$) of independent realizations of the initial thermalized state
[while for the zero-energy initial state, with $T_0=0$, the averaging can be performed over
significantly lower number ($\sim 10^3$) of independent realizations of the initial state].

We consider first the chain with the periodic interatomic potential (\ref{f4}). The left end
of the chain is thermalized at $T_l=1$, and the rest of the chain -- at $T_0$,
when $0\le T_0<T_l$. As one can see in Fig. \ref{fig02}, the thermal energy distribution
substantially depends on the chain temperature $T_0$. For $T_0=0$, the thermal energy
propagates ballistically, with the sound speed,  along the chain. While for $T_0=0.15$,
only relatively small part of thermal energy propagates ballistically, which then
dissipates in the chain. The main part of thermal energy is concentrated at the left,
high-temperature, end of the chain and spreads slowly towards the right end.
For the higher temperature of the chain, $T_0=0.2$, there is no ballistic energy
propagation along the thermalized chain: the energy distribution spreads diffusively from the very beginning.

The unilateral spreading can be described by the following mean square
displacement of thermal energy distribution (MSDTED) of the excess energy, initially placed
in the site $n=1$ of the chain with $u_0\equiv 0$, $u_N\equiv 0$:
\begin{equation}
\langle\Delta x^2\rangle(t) =\sum_{n=1}^{N-1}(n-1)^2e_n(t). \label{f10}
\end{equation}
Here $e_n(t)=(E_n(t)-E_0)/E$ is a discrete distribution of normalized excess energy
in the chain,
$E=\sum_{n=1}^{N-1}(E_n-E_0)$ is a constant total excess energy,
$E_0=T_0$ is the average particle energy in the thermalized chain at temperature $T_0$, see also Refs. \cite{li2014,skc12}.
Mean square displacement can also be introduced for the discrete temperature distribution along the chain, when
$T_n(t) =\langle\dot{x}_n^2(t)$, and both mean square displacements, of thermal energy
and temperature distributions, have the same time dependence.
\begin{figure}[tb]
\begin{center}
\includegraphics[angle=0, width=1\linewidth]{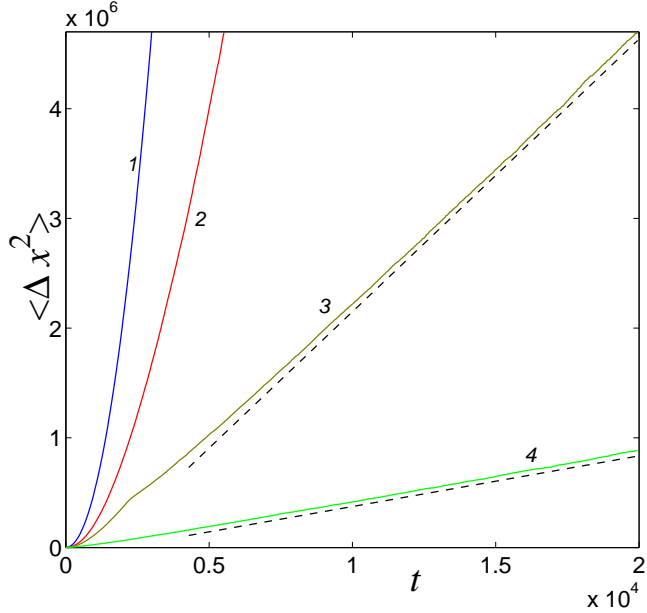}
\end{center}
\caption{
Time dependence of MSDTED,
Eq. (\ref{f10}), in the chain with the periodic potential (\ref{f4}) for the chain
temperature $T_0=0$, 0.15, 0.2, and 0.3 (curves 1, 2, 3, and 4). Dotted lines show the linear dependencies.
}
\label{fig03}
\end{figure}

Time dependence of MSDTED in the chain with the periodic potential (4) is shown
in Fig. \ref{fig03}. As one can see in this figure, at zero and low chain temperature,  $T_0=0$
and $T_0=0.15$, MSDTED grows quadratically
with time, $\langle\Delta x^2\rangle\propto t^2$, which corresponds to the ballistic energy
propagation. At higher temperature $T_0=0.2$, during the short delay time $t< 10^3$
MSDTED also grows quadratically with time, but for longer delay
$t>10^4$ MSDTED changes to become linear in time,
which corresponds to the normal energy diffusion. As one can also see in Fig. \ref{fig03},
the normal energy diffusion starts earlier for the higher temperature of the chain which is a consequence of the enhanced phonon scattering by thermally-activated  anharmonicity of the chain.
From Fig. \ref{fig03} follows that the used chain length $N$ is not enough long to model the transition from the
ballistic to normal diffusion regime of energy propagation at temperature $T_0=0.15$.
We also note that this length of the chain is not enough to model the normal TC in it
at this temperature as well.  We
can conclude from the comparison of Figs. \ref{fig01} and \ref{fig03} that the minimal chain length $N_{min}$, at which the normal TC is established,
is directly  related  with the minimal delay time $t_{min}$, which is needed for the normal energy diffusion
to be established in the same chain: $N_{min}\approx v_st_{min}/a$.
\begin{figure}[tb]
\begin{center}
\includegraphics[angle=0, width=1.\linewidth]{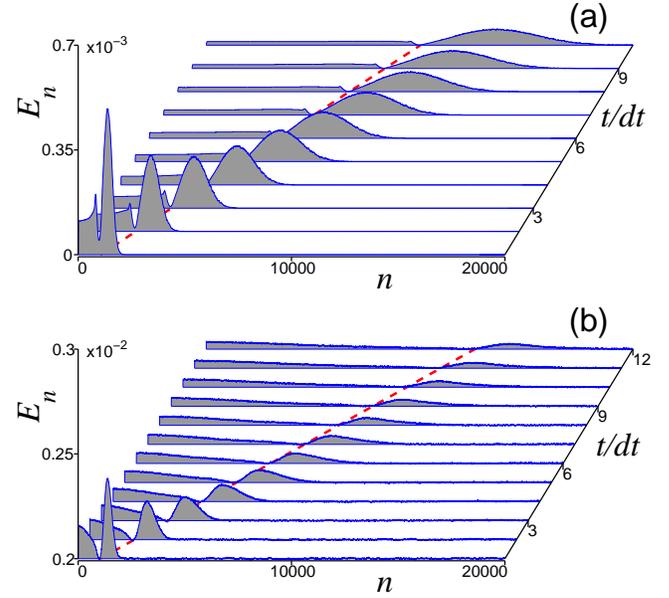}
\end{center}
\caption{
Propagation of energy  $E_n$ in the chain with the Lennard-Jones potential (\ref{f6})
for $T_l=0.01$, $T_0=0.0$ (a) and $T_0=0.002$ (b).
Energy distributions are shown for characteristic delay time $dt=1000$,
dashed lines show sound cone $n=v_st/a$.
}
\label{fig04}
\end{figure}

\begin{figure}[tb]
\begin{center}
\includegraphics[angle=0, width=1\linewidth]{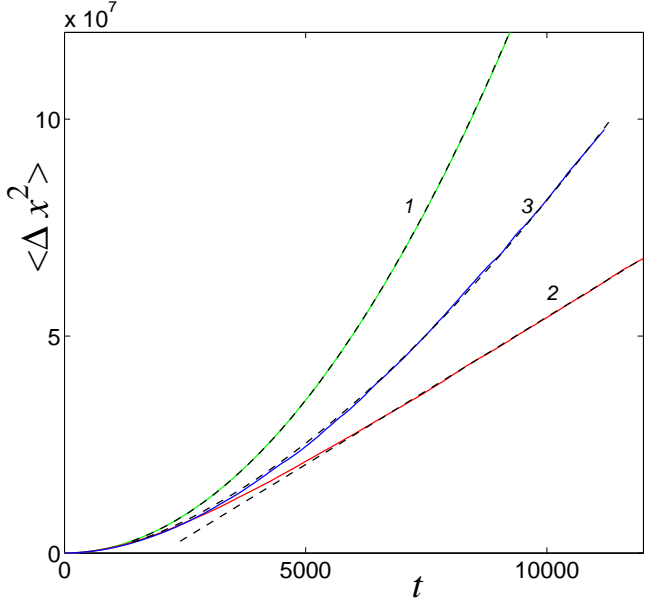}
\end{center}
\caption{
Time dependence of MSDTED,
Eq. (\ref{f10}), in the chain with the Lennard-Jones potential (\ref{f6})
with temperature $T_l=0.01$ of the
left end and temperature
$T_0=0$ and $T_0=0.002$ (solid lines 1 and 2) of the rest of the chain,
and in the chain with the combined potential (\ref{f7})
for $T_0=0.002$, $T_1=0.01$ (solid line 3). Dashed lines 1, 2 and 3
give the quadratic $1.41t^2$, linear $6800(t-2000)$ and power-law $31(t-300)^{1.61}$ dependencies.
}
\label{fig05}
\end{figure}

Similar time dependence of MSDTED is revealed
in the chain with the LJ interatomic potential, see Figs. \ref{fig04} and \ref{fig05}.
We consider the chain with the same $N=20000$, when the first $N_l=40$ atoms in its
left end are thermalized at $T_l=0.01$. As one can see in Fig. \ref{fig04} (a), for $T_0=0$
the energy spreading can be separated into the two parts, which propagate with the sound and supersonic
speeds, respectively.
The first part corresponds to energy transfer by small-amplitude linear lattice waves (phonons)
while the second one corresponds to energy transfer by supersonic kinks (compression acoustic solitons)
produced by hard compression component of the LJ interatomic potential \cite{yak04,archi15,b1}.
For $T_0=0$ MSDTED increases
quadratically with time, $\langle\Delta x^2\rangle\sim t^2$, see line 1 in Fig. \ref{fig05}, which
corresponds to the ballistic energy propagation in the zero-temperature (nonthermalized) chain.

\begin{figure}[tb]
\begin{center}
\includegraphics[angle=0, width=1\linewidth]{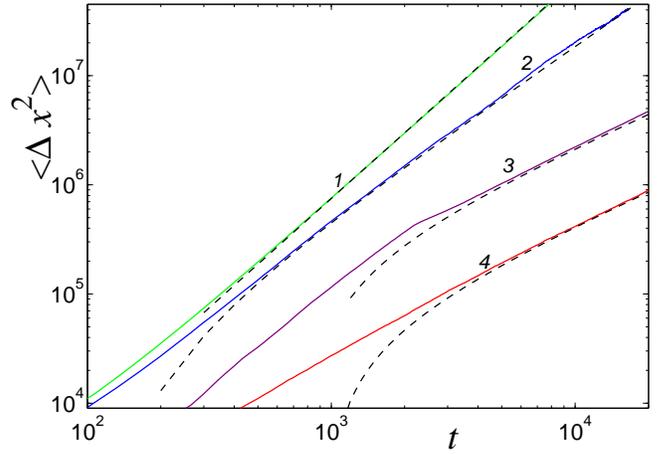}
\end{center}
\caption{
Time dependence of MSDTED,
Eq. (\ref{f10}), in the chain with the combined potential (\ref{f5})
with temperature $T_l=1$ of the left end and  temperature
$T_0=0$ and $T_0=0.3$ (solid lines 1 and 2) of the rest of the chain.
Dotted lines 1 and 2 give the dependencies $0.7t^2$ and
$12.5(t-110)^{1.544}$. Solid lines 3 and 4 give the dependencies for the chain with the periodic
interatomic potential (\ref{f4}) for $T_0=0.2$ and $T_0=0.3$.
Dashed lines 3 and 4 give the dependencies $230(t-800)$ and $45(t-964)$.
}
\label{fig06}
\end{figure}

The picture of energy propagation changes in the thermalized chain. For $T_0=0.002$,
the part of the initial energy, which propagates with supersonic speed, starts to dissipate
in the chain (because the acoustic solitons have finite mean free path in the thermalized
LJ chain), and the normal diffusion of thermal energy is established in the chain, see Fig. \ref{fig04} (b).
Time dependence of MSDTED  shows that for $t>5000$ the initial ballistic energy propagation
(superdiffusion) is replaced by the normal energy diffusion
when $\langle\Delta x^2\rangle$ grows linearly with time, see line 2 in Fig. \ref{fig05}.
In the chain with the combined potential (\ref{f5}) or (\ref{f7}),
the anomalous superdiffusion of energy takes place and MSDTED grows as a power function of time: $\langle\Delta x^2\rangle\propto t^\beta$
with the exponent $1<\beta<2$. In the chain with potential (\ref{f7}), $\beta=1.61$ at $T_0=0.002$, see line 3 in Fig. \ref{fig05}; in the chain with potential (\ref{f5}), $\beta=1.544$ at $T_0=0.3$, see line 2 in Fig. \ref{fig06}. Again the minimal delay time, at which the normal energy diffusion is established, is consistent
with the minimal chain length, at which the normal TC is established, cf. Figs. \ref{fig05} and \ref{fig06}  with Fig. \ref{fig01}.

The value of the exponent $\beta$ can be related with the exponent  $\alpha$ in the length dependence
of anomalous TC, when $\kappa\propto N^\alpha$.
By the definition of the thermal
energy diffusion $D_E$ and TC $\kappa=cD_E$ coefficients, where $c$ is the specific heat density, MSDTED grows as a power function of time as
\begin{eqnarray}
\label{f11}
\langle\Delta x^2\rangle &\propto& D_E(N)t\propto N^\alpha t, \quad \\
\langle\Delta x^2\rangle &\propto& t^\beta, \quad
\label{f12}
\end{eqnarray}
where $D_E(N)=\kappa(N)/c\propto N^\alpha$. Below we show that (at least) two different relations between $\alpha$ and $\beta$
can be obtained from Eqs. (\ref{f11}) and (\ref{f12}) under different assumptions.
According to its definition (\ref{f10}),
$\langle\Delta x^2\rangle$ scales as the square of the dimensionless length.
Then under the assumption
\begin{equation}
\langle\Delta x^2\rangle \propto N^2,
\label{f13}
\end{equation}
we get from Eqs. (\ref{f11}) and (\ref{f12}) that
\begin{equation}
\alpha=2-2/\beta. \label{f14}
\end{equation}

Relation (\ref{f13}) corresponds to the assumption that the regime of the thermal diffusion
is reached only in the chain, whose length reaches (or exceeds) the effective phonon mean free path.
But under the different assumption
\begin{equation}
t \propto Na/v_s, \label{f15}
\end{equation}
we get from Eqs. (\ref{f11}) and (\ref{f12}) that
\begin{equation}
\alpha=\beta-1. \label{f16}
\end{equation}

Relation (\ref{f15}) corresponds to the assumption that the energy carriers launched from the hotter
side of the chain will (ballistically) reach the colder side and reflect back beyond the delay  time $t$.
Both scaling relations (\ref{f14}) and (\ref{f16}) imply that normal energy diffusion ($\beta=1$)
leads to the normal (non-divergent) TC ($\alpha=0$),
while the superdiffusion ($\beta>1$) corresponds to anomalous ($\alpha>0$) TC.
Relation (\ref{f14}) was suggested in Refs. \cite{lw03,l05}, while the relation (\ref{f16})
was obtained in Refs. \cite{den2003,cdp05,dhar2013,li2014} in the specific case of billiard-like 1D
models in which noninteracting particles undergo L\'{e}vy flights.
It is worth noting that the assumption (\ref{f13}) was implicitly used  in derivation of Eq.~(\ref{f14})
in Ref.~\cite{lw03}, see also Ref.~\cite{metz04}, and the assumption (\ref{f15}) was explicitly used
in derivation of Eq.~(\ref{f16}) in Ref.~\cite{li2014}. Moreover, the scaling relation (\ref{f14})
can be derived from the analysis of the diffusion with a position-dependent diffusion coefficient, see, e.g., Ref. \cite{sok00},  under the same assumption (\ref{f13}).

The main conclusion of our analysis is that neither of the relations (\ref{f14}) and (\ref{f16})
is the universal relation, which can be applied to all the nonlinear systems with anomalous
heat transport and superdiffusion of thermal energy. We can compare this conclusion with the known conjecture
on the absence of the unique velocity-correlation function in turbulent flow, which is universal
for all relevant scales and types of flow, see  Refs. \cite{llhydro,frisch}.
Our modeling of anomalous thermal conductivity and superdiffusion of thermal energy in
the chains with the combined interatomic potentials (\ref{f5}) and (\ref{f7})
confirms with high accuracy the relation (\ref{f14}): we have $\alpha=0.705$, $\beta=1.544$
for the chain with the potential (\ref{f5}), and $\alpha=0.76$, $\beta=1.61$ for the chain with
the potential (\ref{f7}). On the other hand, the relation (\ref{f16}) was confirmed in the study
of heat transport in the billiard-like 1D system containing
colliding particles with two different masses at $T_0=0$ \cite{cdp05,szg15}.
We relate the difference between Eqs. (\ref{f14}) and (\ref{f16}) with the fact
that Eq. (\ref{f14}) is applied mostly to the 1D lattices of the coupled (anharmonic)
oscillators, in which heat is transported by  weakly-scattered $waves$ (phonons),
while Eq. (\ref{f16}) is applied mostly to the billiard-like 1D systems,
in which heat is transported by noninteracting $particles$ performing L\'{e}vy flights.

In conclusion, we show that the normal thermal conductivity is always accompanied by the normal
energy diffusion in the thermalized anharmonic chains, while the superdiffusion of energy
is inherent in the thermalized chains with only anomalous heat transport. We
confirm that the confining interparticle potential  makes both heat transport and energy
diffusion anomalous in low-dimensional phononic systems. We show that the scaling relation
between the exponents in time dependence of the mean square displacement
of thermal energy distribution and in length dependence of anomalous thermal conductivity
is not universal and can be different, depending on the main mechanism of energy transport:
either by weakly-scattered waves or by noninteracting colliding particles performing L\'{e}vy flights.

The authors are grateful to the Joint Supercomputer Center
of the Russian Academy of Sciences for the use of computer facilities.

\end{document}